\documentclass{article}
\usepackage[margin=2.9cm]{geometry}
\usepackage[pdftex]{graphicx}
\usepackage{amsmath}
\usepackage{hyperref}
\usepackage{fancyhdr}
\begin{document}
\fancyhf{}
\lhead{O'Malley-James \textit{et al.}}
\rhead{Swansong Biospheres}
\cfoot{\thepage}

\begin{titlepage}

\Large
\begin{center}
Swansong Biospheres: Refuges for life and novel microbial biospheres on terrestrial planets near the end of their habitable lifetimes
\end{center}
~

\normalsize
\noindent Jack T. O'Malley-James
\newline
School of Physics and Astronomy, University of St Andrews, North Haugh, St Andrews, Fife, UK.
\newline
~
\newline

\noindent Jane S. Greaves
\newline
School of Physics and Astronomy, University of St Andrews, North Haugh, St Andrews, Fife, UK.
\newline
~
\newline

\noindent John A. Raven
\newline
Division of Plant Sciences, University of Dundee at TJHI, The James Hutton Institute, Invergowrie, Dundee, UK.
\newline
~
\newline

\noindent Charles S. Cockell
\newline
UK Centre for Astrobiology, School of Physics and Astronomy, James Clerk Maxwell Building, The King’s Buildings, University of
Edinburgh, Edinburgh, UK.
\newline
~
\newline
~
\newline
~
\newline
~
\newline

\noindent\textbf{Corresponding author:}
\newline
\noindent J.T. O'Malley-James
\newline
School of Physics \& Astronomy
\newline
University of St Andrews
\newline
North Haugh
\newline
St Andrews
\newline
Fife, KY16 9SS
\newline
E.mail: \url{jto5@st-andrews.ac.uk}

\end{titlepage}

\begin{center}
\noindent\textbf{Abstract}
\end{center}

\noindent\textit{The future biosphere on Earth (as with its past) will be made up predominantly of unicellular microorganisms. Unicellular life was probably present for at least 2.5 Gyr before multicellular life appeared and will likely be the only form of life capable of surviving on the planet in the far future, when the ageing Sun causes environmental conditions to become more hostile to more complex forms of life. Therefore, it is statistically more likely that habitable Earth-like exoplanets we discover will be at a stage in their habitable lifetime more conducive to supporting unicellular, rather than multicellular life. The end stage of habitability on Earth is the focus of this work. A simple, latitude-based climate model incorporating eccentricity and obliquity variations is used as a guide to the temperature evolution of the Earth over the next 3 Gyr. This allows inferences to be made about potential refuges for life, particularly in mountains and cold-trap (ice) caves and what forms of life could live in these environments. Results suggest that in high latitude regions, unicellular life could persist for up to 2.8 Gyr from present. This begins to answer the question of how the habitability of Earth will evolve at local scales alongside the Sun's main sequence evolution and, by extension, how the habitability of Earth-like planets would evolve over time with their own host stars.}
\newline

\noindent\textbf{INTRODUCTION}
\newline

The prospects of finding a habitable zone terrestrial planet are looking increasingly good. The fraction of Sun-like stars with Earth-like planets in their habitable zones ($\eta_{\oplus}$) is currently estimated to be between 1.4-2.7\% (Catanzarite \& Shao, 2011) and there are tentative suggestions that $\eta$$_{\oplus}$ could be as high as 42\% for M star planet hosts (Bonfils \textit{et al.}, 2011). While knowing the conditions that make a planet habitable is clearly extremely useful when searching for life, it is equally useful, although perhaps more pessimistic, to ask the question: when will a planet cease to be habitable? It is entirely possible that some future discoveries of habitable exoplanets will be planets that are nearing the end of their habitable lifetimes, i.e. with host stars nearing the end of their main sequence lifetimes. 

\begin{figure}
		\includegraphics[scale=0.85]{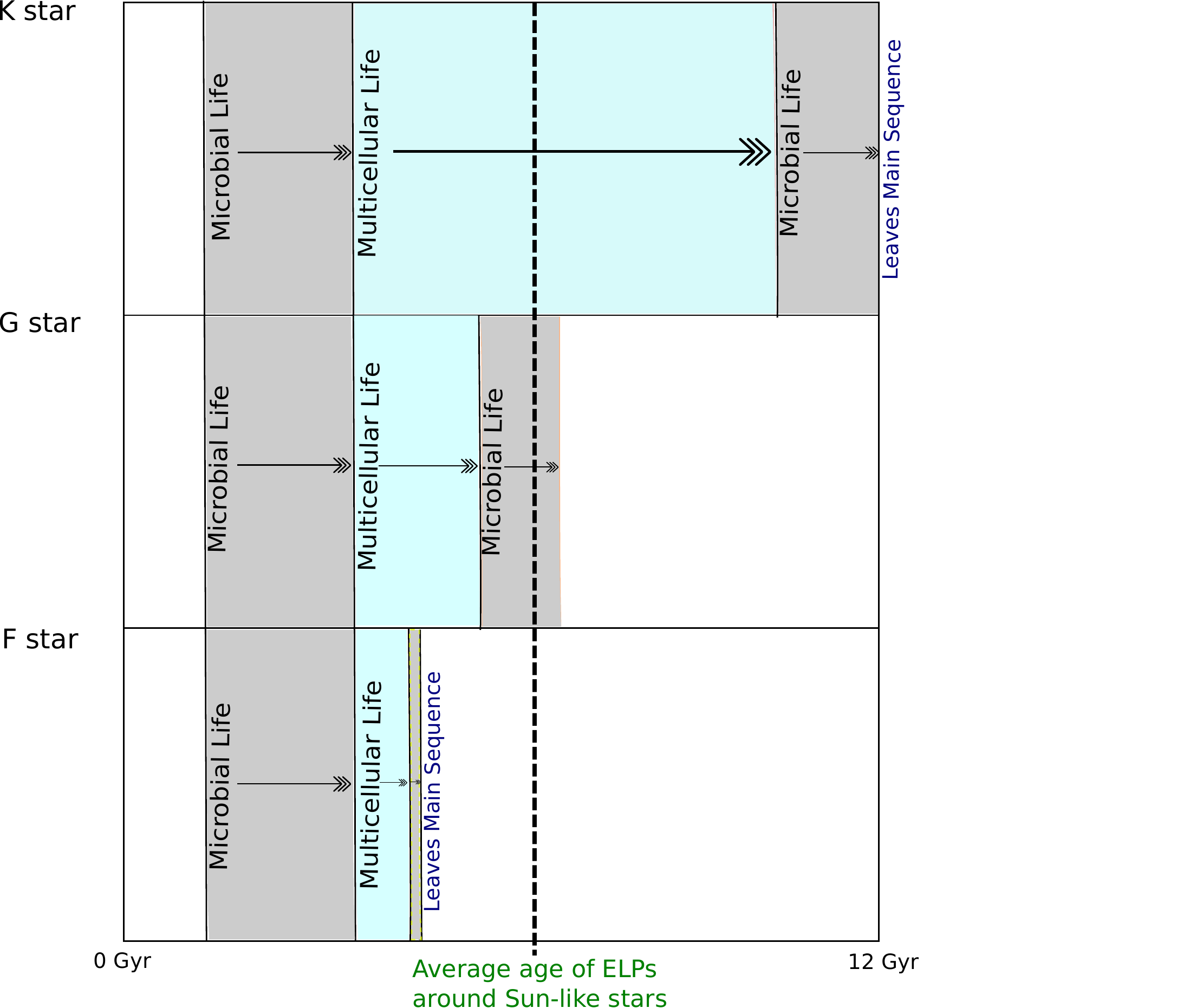}
           \caption{\textit{Time windows for complex and microbial life on Earth analogue planets orbiting Sun-like stars (F(7), G and K(1) stars) during their main sequence lifetimes. Assuming that the processes leading to multicellular life are the same as on Earth (i.e. $\sim$1 Gyr for life to emerge and $\sim$3 Gyr for multicellularity to evolve), the potential lifespans of a more complex, multicellular biosphere are estimated. Multicellular life was assumed able to persist until surface temperatures reach the moist greenhouse limit for an Earth analogue planet in the continuously habitable zone of the star-type in question. Microbial life is then assumed to dominate until either the maximum temperature for microbial life is exceeded, or until the star leaves the main sequence (whichever happens first). The average age of Earth-like planets was found by Lineweaver (2001) to be $\sim 6.4 \pm 0.9$ Gyr based on estimates of the age distribution of terrestrial planets in the universe. This average age falls within microbial and uninhabitable stages for G and F type stars respectively, but falls within the multicellular life stage for K stars.}\label{Fig:fig1}}
\end{figure}

\begin{figure}
                    \centering
		\includegraphics[scale=0.65]{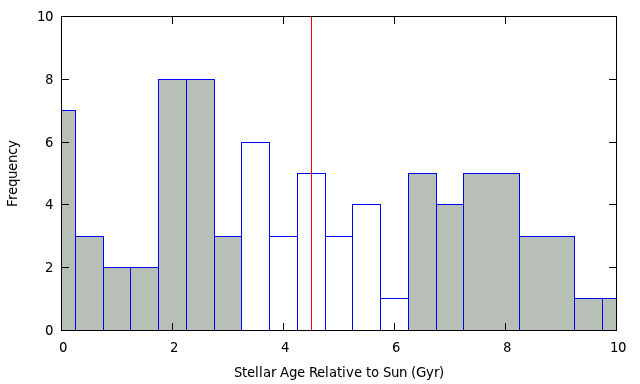}
           \caption{\textit{Main sequence evolution (relative to the Sun and accounting for stellar type) of currently known terrestrial planet (10$M_E$ and less) hosting stars. The left shaded region represents early microbial biospheres and the right shaded region represents late microbial biospheres for Earth analogue planets. The central unshaded region represents the stellar age-range within which Earth analogue planets would be more likely to have complex biospheres. The vertical line represents the current age of the Sun. Source: \textit{The Extrasolar Planets Encyclopaedia} (Schneider, 2010).}\label{Fig:fig2}}
\end{figure}

An impression of the stages a habitable Earth analogue planet (here taken to be an exact copy of Earth) passes through is given in \textit{Figure \ref{Fig:fig1}} while the fraction of known terrestrial planets around late main sequence stars is
\noindent illustrated in \textit{Figure \ref{Fig:fig2}} . Knowing the likely conditions on a far-future Earth analogue will help to determine whether any life still present on such a planet will be remotely detectable. Due to the harsh environmental conditions such planets would face, it is expected that this life would be microbial.
\newline
\indent  Life is thought to have emerged on Earth 3.8 Gyr ago (Schidlowski, 1988; Rosing, 1999) - and perhaps even as early 3.85 Gyr ago (Mojzsis \textit{et al.}, 1996). Chemical evidence suggests that unicellular organisms were present for at least 2.5 Gyr before body fossil evidence of multicellular life appeared, as the red alga \textit{Bangiomorpha} 1.2 Gyr ago (Butterfield, 2000; Strother \textit{et al.}, 2011), while most animal phyla (groups of organisms with a degree of evolutionary relatedness) did not appear as body fossils until the `Cambrian Explosion\rq{} 530 Myr ago. Due to their greater metabolic and environmental versatility and ability to survive - including in a dormant state - under extreme physical and chemical conditions, microbes are likely to be the only forms of life capable of surviving in the hostile environments that will be found on the far-future Earth. They are, by many criteria, the most diverse, abundant and successful forms of life on Earth, notable for not only surviving, but often thriving in what, from an anthropocentric view point, would be considered extreme environments (Cockell, 2003). Hence, it is reasonable to assume that microbial life will be the most abundant form of life (if it is found) in habitable extra-terrestrial environments. In fact, their ability to survive in extreme environments (such as the microbes living at temperatures of 90$^{\circ}$C in Grand Prismatic Spring in Yellowstone National Park, or those that live in the highly acidic waters of the Rio Tinto, Spain (Dartnell, 2011)) opens up the range of extra-terrestrial environments that could be classified as habitable.
\newline
\indent On Earth such organisms exhibit more modes of life beyond the mechanisms of oxygenic photosynthesis and aerobic respiration used by ``higher\rq{}\rq{} plants (those that are most structurally divergent in differentiation among their cells) and animals (Oren, 2009). Thus, microbial life produces biosignatures that are unlike those produced by the majority of life on Earth as a whole. For example, Kaltenegger \textit{et al.} (2007) show that the early Earth (a predominantly microbial world) exhibited very different atmospheric biosignatures to the present Earth. Potentially, remotely detected biosignatures from an exo-Earth with just a microbial biosphere may not be immediately considered to be caused by life due to the dissimilarities to the signatures of life on the present Earth. This work is a first step towards simulating these more novel microbial environments under a diverse range of radiation regimes and surface environmental conditions to ascertain the likely remotely detectable biosignatures that would be produced.
\newline
\indent The host planets considered in these simulations will often not be direct Earth analogues, and given galactic statistics, they will often be found around red dwarf stars (Tarter \textit{et al.}, 2007) and in binary star-systems (Welsh \textit{et al.}, 2012). Each host star system will provide a unique set of planetary environmental conditions from varying irradiation and stellar flare activity to unusual surface compositions and temperature ranges.
\newline
\indent When developing  any model of this nature, it is useful to calibrate it using a well known system; hence, initially the Earth itself was used to test the model.
\newline

\noindent\textbf{Microbial Niches on the Far-Future Earth}
\newline
\indent The Sun's main sequence evolution will result in very extreme environments for life on the far-future Earth (taken here to be 2-3 Gyr from the present). As the Sun ages, its luminosity increases, resulting in surface temperature increases on Earth.  Increased temperatures cause increased atmospheric water vapour - a greenhouse gas, the presence of which further increases surface temperatures. However, this simplified overview neglects the role of cloud cover. Clouds can reduce planetary heating by reflecting solar energy away from the planet, increasing the planetary albedo, but can also act to trap radiation, increasing a heating effect (McGuffie \& Henderson-Sellers, 2005). More water vapour entering the atmosphere could lead to increased cloud formation, but there are uncertainties regarding the effect of clouds and greenhouse warming (Goldblatt \& Watson, 2012) as a result of the number of dynamical and thermodynamical factors that would need to be considered (McGuffie \& Henderson-Sellers, 2005). It is beyond the scope of this investigation to model this. Therefore, a constant cloud cover will be assumed in this case. Boer \textit{et al.} (2004) found that cloud cover variations were small until the solar constant increases by 25\% of its present value; hence, this should be a reasonable assumption for most of the modelled time period.
\newline
\indent Higher temperatures lead to increased weathering of silicate rocks, drawing down more carbon from the atmosphere. Carbon is normally recycled though plate tectonics; however, increasing water loss eventually halts plate tectonics due to greater friction between the plates. Lower atmospheric carbon dioxide levels make embryophytic plant (liverworts, hornworts, mosses and vascular plants) life unsustainable as, unless a more effective photosynthetic carbon acquisition mechanism is used (Bar-Even \textit{et al.}, 2010; Bar-Even \textit{et al.}, 2012), plants require a minimum atmospheric CO$_2$ concentration of approximately 10 p.p.m. (Caldeira \& Kasting, 1992). Even with a higher-affinity carboxylase, photosynthesis in air would be restricted by diffusion in the gas phase through the diffusion boundary layer at very low bulk phase CO$_2$ concentrations. Once levels drop below this value higher plants will begin to die off. This in turn decreases oxygen production, which, with continued consumption by biota and by oxidation of kerogen (organic carbon in sedimentary rocks), leads to a steady decline in atmospheric oxygen to zero over a few million years (Walker, 1991). Since multicellular (metazoan) animal life depends on oxygen for respiration, the end of animal life would occur a few million years after the end of plant life (Walker, 1991). Large endotherms (mammals, birds) would likely be the first group to become extinct due to their higher oxygen requirements compared to smaller endotherms and exotherms. As temperatures increase alongside declining atmospheric oxygen levels, eutherians (placental mammals) would be particularly vulnerable. Not only do these have higher oxygen requirements than non-placental mammals, which is said to have delayed the evolution of large placental mammals until atmospheric oxygen reached a certain level (Falkowski \textit{et al.}, 2005), but embryo development is very sensitive to excess heat (McLean, 1991). Large herbivorous mammals would suffer from the decrease in food supplies as plant abundance decreases. Smaller mammals would have a slight stay of execution in comparison, due to their lower oxygen requirements and their larger surface-to-volume quotient, which aids in the dissipation of heat (McLean, 1991). This scenario is slightly complicated by the fact that eutherians have higher body temperatures than marsupials and monotremes (Clarke \& Rothery, 2008), which may, therefore, be more susceptible to increasing temperatures. Birds may be better suited to surviving than mammals larger than the largest birds as their generally smaller sizes mean they require less oxygen than larger animals, but also, migratory birds especially would be better able than similarly sized mammals to travel long distances to find lower temperature refugia; however, the number of these  refugia for such animals would decline as temperatures continue to rise. These refugia would also tend to be at higher elevations where less land surface area is available, restricting population sizes as species migrate upwards (Sekercioglu \textit{et al.}, 2007).

Ectothermic vertebrates (fish, amplibians, reptiles) would be able to survive for longer than endotherms in this scenario due to their better heat tolerances and, in general, lower oxygen requirements (some ectotherms have been observed to consume oxygen in greater quantities than living endotherms of the same body weight) (Kemp, 2006). Decreasing water availability would make some amphibian species more vulnerable in such a world (Ar\'{a}ujo \textit{et al.}, 2006). Fish species would be similarly vulnerable. It should be noted that rapid ocean evaporation would not yet have begun at this time. Marine species may be able to survive for longer than freshwater species due to the orders of magnitude greater volume of ocean water than freshwater. For ectotherms, external temperature influences metabolic rate. An increased environmental temperature results in an increased metabolic rate and therefore, an increased need for food. Hence, surviving species may be vulnerable to starvation (Dillon \textit{et al.}, 2010). Reptile species with temperature-dependent sex determination would be more susceptible to increased temperatures (Ar\'{a}ujo \textit{et al.}, 2006).
\begin{figure}
                    \centering
		\includegraphics[scale=0.65]{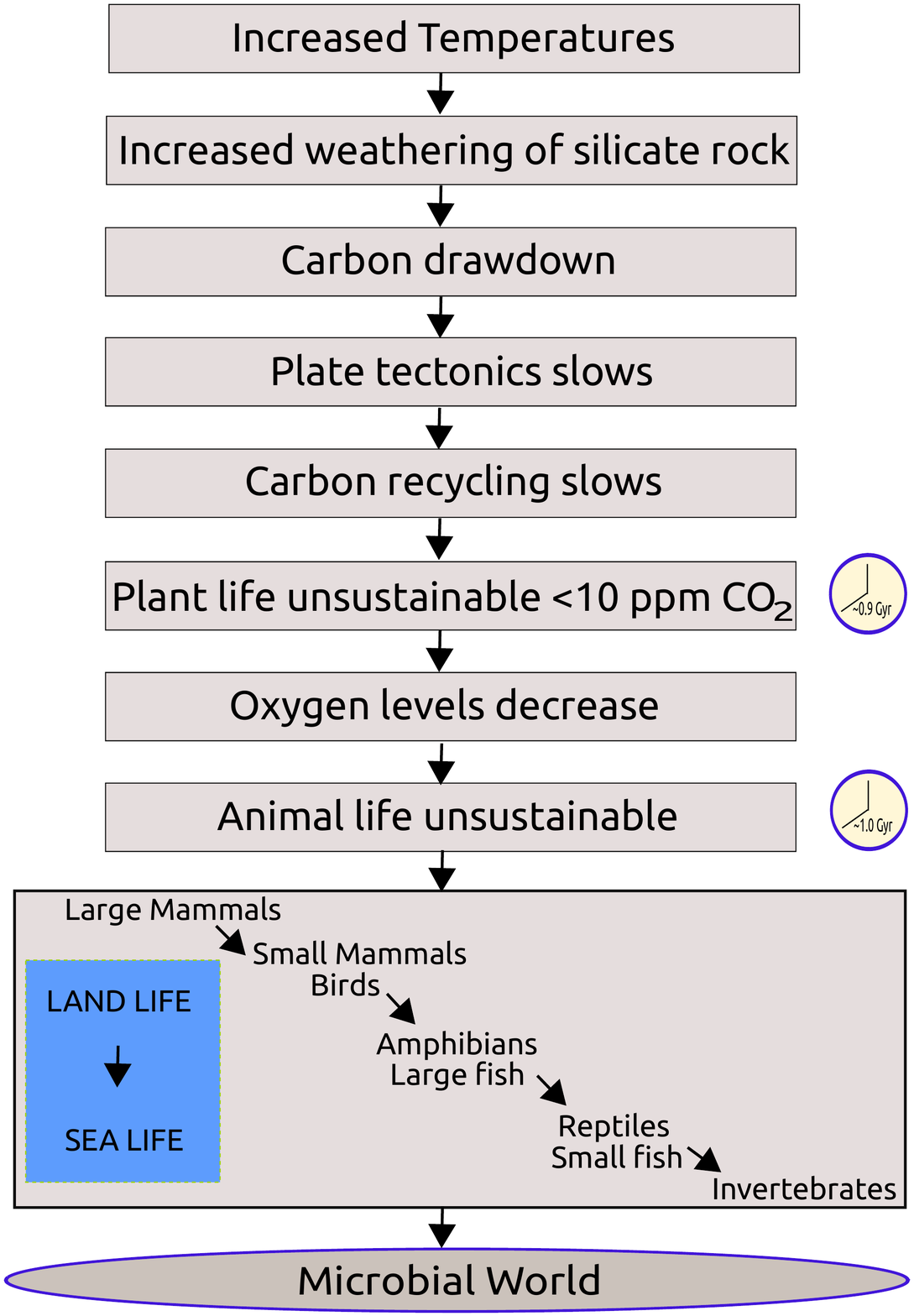}
           \caption{\textit{Simplified sequence of events leading to the loss of plant and animal life on a rapidly warming planet and the resulting microbial world.}\label{Fig:fig3}}
\end{figure}

Invertebrates could be the final animals present on Earth before all animal species are lost. Some insects, such as the spider beetles \textit{Mezium affine} and \textit{Gibbium aequinoctiale} have been observed to survive (but not necessarily to complete their life cycle) in temperatures of up to 56$^{\circ}$C (Yoder \textit{et al.}, 2009). In general, life on land will be more vulnerable than marine life initially as a result of the temperature-buffering effects of water. However, the loss of plant life on land would lead to a reduction in nutrients reaching the ocean. Communities isolated from marine food chains, such as volcanic vent communities, would likely survive for the longest (Ward \& Brownlee, 2002). Within approximately 0.1 Gyr after the end of higher (embryophytic) plant life, Earth will once more become a microbial world.

\begin{figure}
                    \centering
		\includegraphics[scale=0.55]{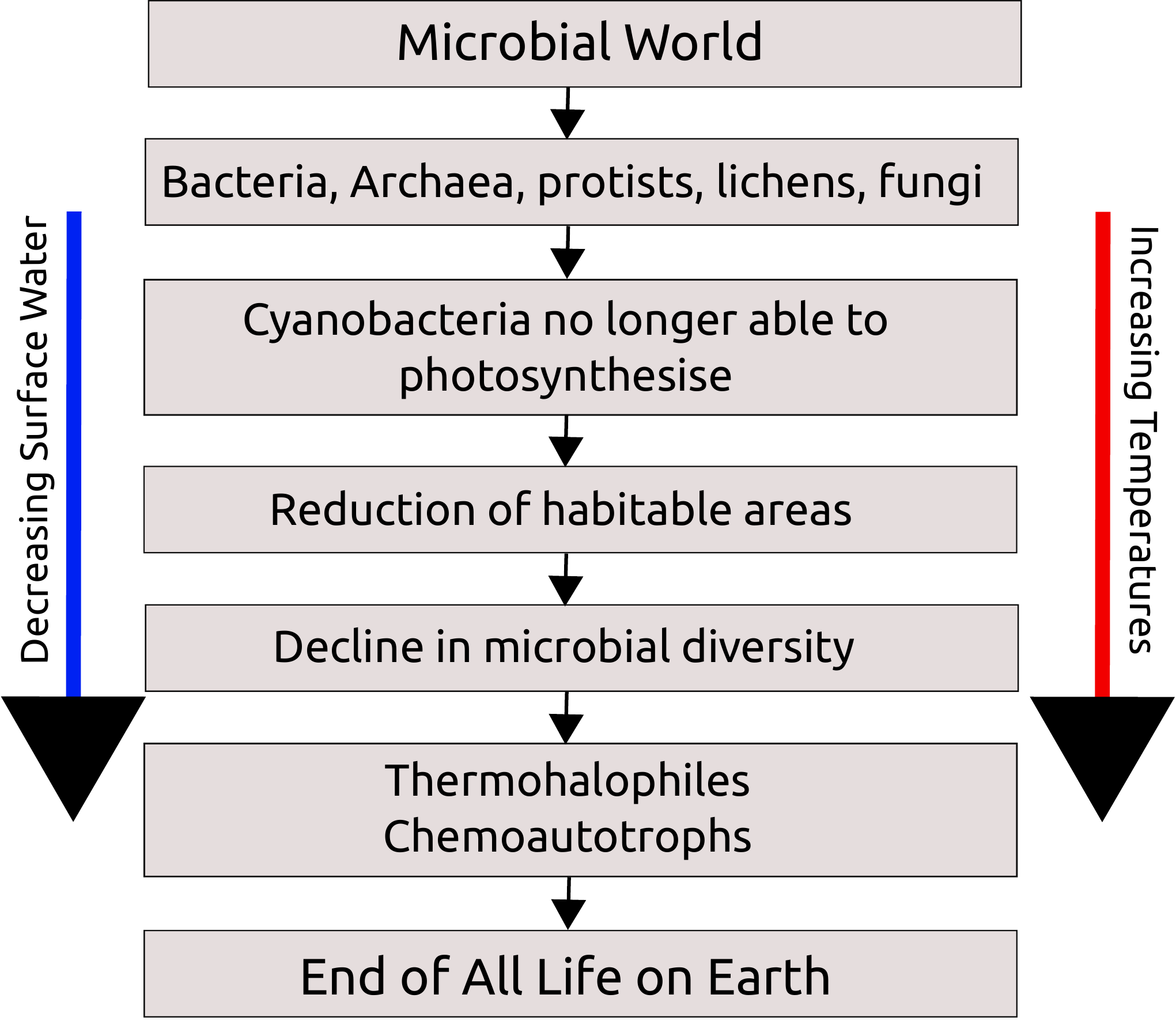}
           \caption{\textit{Simplified sequence of events leading from a microbial world with increasing temperatures and decreasing water availability to the end of all life on Earth.}\label{Fig:fig4}}
\end{figure}

Microbial photosynthesis could continue for a further 0.1 Gyr. Some eukaryotic pytoplankton and benthic macroalgae, and cyanobacteria, can grow and persist in environments with large CO$_2$ fluctuations (Maberly, 1996) and can carry out net photosynthesis at very low CO$_2$ concentrations (Birmingham \& Colman, 1979; Maberly, 1990). Hence, carbon dioxide concentrating mechanisms in cyanobacteria and, to a smaller extent, some eukaryotic algae could allow photosynthesis to take place down to carbon dioxide concentrations of 1 ppm; 10 times less than the minimum for terrestrial higher (C4 photosynthesis) plant life (Caldeira and Kasting 1992). A steady decline in habitable environments with time leads to a pruning of the tree of life, concentrated towards extremophilic life adapted to multiple extreme conditions (polyextremophiles (Mesbah \& Wiegel, 2012)), such as thermohalophiles and chemoautotrophs. Finally, rising temperatures cause rapid ocean evaporation (Kasting, 1988; Bounama \textit{et al.}, 2001; V\'{a}squez et al., 2010), effectively sterilising most of the biosphere (see \textit{Figures \ref{Fig:fig3}} and \ref{Fig:fig4}).

It should be noted that it may be possible that the increase in atmospheric oxygen caused by the photodissociation of an ocean’s worth of water vapour could cause the extinction of those final lifeforms that cannot tolerate high oxygen concentrations (similar to the extinction believed to have been caused by the Great Oxidation Event 2.4 Gyr ago) even before all the remaining pools of liquid surface water are lost. However, this future oxygen atmosphere would only persist for a geologically short period of time before oxygen is removed via oxidation, and surface reactions. Some oxygenic photosynthetic organisms are able to survive and grow at oxygen equilibrium gas phase partial pressures of 80 kPa in solution (at sea level) (Raven \& Larkum, 2007); therefore, organisms with similar tolerances would be better suited to surviving a brief high-oxygen period.
With the complete cessation of photosynthesis, but the continuation of all other biological activity at the present-day rate, the half-life of atmospheric oxygen would be approximately 3 Myr (Walker, 1991); hence, the geologically short timespan for the continuation of animal life after higher plant life comes to an end.

On the the far-future Earth, the only likely oxygen sink would be nitrogen fixation by lightning. At present there is approximately $6\times 10^{19}$ moles of atmospheric oxygen in the atmosphere. With $5\times 10^{11}$ moles of nitrogen converted to nitric acid by lightning each year (Tie \textit{et al.}, 2002), atmospheric oxygen would have a half-life of approximately 100 Myr if fixation by lightning were the only oxygen removal process. However, lightning frequency is directly linked to a planet's climate, with the frequency of lightning strikes increasing with increased surface temperature (Williams \textit{et al.}, 2005; Price \& Asfur, 2006a, Sekiguchi \textit{et al.}, 2006). Lightning frequency has also been observed to increase with increased cloud cover in the tropics (Sato \& Fukunishi, 2005). Lightning modelling studies suggest that global lightning activity increases by 10\% for every 1 K increase in global temperature (Price, 2008). Hence, the rate of atmospheric oxygen removal on a warmer, far-future Earth would be much higher than present rates. For example, a global temperature 100 K higher than that at present would result in approximately $6\times 10^{15}$ mols of nitrogen converted into nitric acid per year. Even with a higher level of atmospheric oxygen than the present level (the ocean contains approximately $7\times 10^{22}$ mols of oxygen), the half-life of atmospheric oxygen would be of the order 6 Myr, suggesting a geologically short oxygen-rich period during the rapid photodissociation period. However, the question of whether this would be a planet-sterilising event remains open.

As the climate evolves alongside solar main sequence evolution, life on Earth and its associated biosignatures will also evolve until the point at which the planet becomes uninhabitable. This raises the question of when biological activity on Earth would be reduced to such an extent that the planet no longer exhibits remotely detectable biosignatures.

Assuming that remotely detectable life resides on, or near the surface (signatures of deep biospheres tend to require direct sampling of the biosphere itself (Parnell \textit{et al.}, 2010)) and that such life requires liquid water (Rothschild \& Mancinelli, 2001), a step towards answering this question was taken by constructing a model to find the most likely regions on the far-future Earth for the planet's ``last'' life.

While solar evolution will ultimately be responsible for the expiration of the biosphere, there are many other factors that will effect the climate evolution on Earth along the way. The orbital parameters of the planet, specifically its eccentricity and obliquity, are known to vary over time, resulting in climate changes as they do so (Spiegel \textit{et al.}, 2010). Atmospheric composition also has a role to play. Changing concentrations of greenhouse gases in particular have a notable effect on surface temperatures (Kasting \& Grinspoon, 1991; Pavlov \textit{et al.}, 2000). Climate changes themselves can then alter other climate-influencing factors such as planetary albedo (Rosing \textit{et al.}, 2010). These factors were taken into account alongside an evolving Sun to create a toy latitude-dependent radiative balance model for Earth to better constrain where the last habitable areas on the planet may be.
\newline

\noindent\textbf{METHODS - Factors Influencing Habitability}
\newline
\noindent (i) Solar luminosity
\newline
\indent The primary driver for habitability change is the increase in solar luminosity over time. This can be accounted for by following the convention of Gough (1981) in which the luminosity at any time during the Sun\rq{}s main sequence lifetime $L(t)$ is given by:
\begin{equation}
L(t) = \left[ 1 + \frac{2}{5}\left(1-\frac{t}{t_{\odot}}\right) \right]^{-1}L_{\odot}
\end{equation}
\noindent where $L_{\odot}$ is the present solar luminosity, $t_{\odot}$ is the current age of the Sun and $t$ is the time elapsed on the main sequence. The surface insolation depends on the solar constant $S_0$, which is influenced by solar luminosity such that
\begin{equation}
S_0 = \frac{L_{\odot}}{4\pi kd_E^{2}}
\end{equation}
\noindent where $k$ is a constant ($k \approx 1$) and $d^2_E$ is the Earth-Sun distance. By incorporating a varying luminosity and assuming a total insolation $S_T = \frac{S_0}{4}$ (as the surface of a sphere has an area four times that of a flat disk with the same radius), the effect of increased luminosity on mean global temperatures can be modelled.
\newline
\indent The time evolution of this mean temperature can be calculated from the difference between incoming and outgoing radiation, i.e.
\begin{equation}
T_{t+1} = \left(\frac{F_{in} - F_{out}}{C_p}\right)\Delta t + T_t
\end{equation}
where $F_{in}$ and $F_{out}$ are the incoming and outgoing radiation fluxes and $C_p$ is the heat capacity of the planet (at constant pressure). In this case, the incoming radiation is assumed to be a function of the solar insolation
\begin{equation}
F_{in}=(1-a)S_T
\end{equation}
\noindent where $a$ is the temperature-dependent albedo and the outgoing radiation is given by
\begin{equation}
F_{out}=(-3g_{atm}\sigma T_0^4) + (4g_{atm}\sigma T_0^3)T(t)
\end{equation}
\noindent where $g_{atm}$ is a variable factor accounting for the transmissivity of the atmosphere ($0 \le g_{atm} \le 1$), $\sigma$ is the Stefan-Boltzmann constant and $T_0$ is the initial temperature of the system. Heat is then allowed to diffuse out from the equator to higher latitudes as described in Lorenz (2001), i.e. representing the convective heat flow from the tropics (a hot reservoir) to mid-latitudes and the poles (cold reservoirs). The magnitude of this heat flow is controlled by the latitudinal diffusion coefficient, which increases as temperatures rise and cause an increase in atmospheric density (Lorenz \textit{et al.}, 2001).
\newline

\noindent (ii) Greenhouse gases
\newline
\indent The levels of greenhouse gases in the atmosphere affect the outgoing radiation lost to space and, therefore a planet\rq{}s surface temperature. Atmospheric gases such as carbon dioxide, water vapour, methane etc. absorb and emit radiation at infrared wavelengths contributing to a global greenhouse effect. The change in longwave radiation forcing due to CO$_2$ and water vapour is proportional to the logarithm of their concentration in the atmosphere (Myhre \textit{et al.}, 1998; R\'{a}k\'{o}czi \& Iv\'{a}nyi, 1999; Allan, 2011).
\newline
\indent In this model, particular attention is paid to the levels of water vapour in the atmosphere (a result of rapid ocean evaporation caused by a runaway greenhouse effect) and the effects of carbon draw-down (the removal of atmospheric carbon as a result of the weathering of silicate rocks).
\newline
\indent This is included in the model by taking account of the partial optical thickness of the atmosphere due to both CO$_2$ and H$_2$O. This gives a total longwave optical depth ($\tau$) made up of the partial optical depths of CO$_2$ and H$_2$O of
\begin{equation}
\tau = k_{CO_2}P_{CO_2}^{0.53}+k_{H_2O}P_{H_2O}^{0.3}
\end{equation}
(Levenson, 2011) where $k_i$ is a proportionality constant and $P_i$ is the partial pressure, which evolves as the number density ($n_i$) evolves with time such that
\begin{equation}
P_i=\frac{n_iRT}{N_A}
\end{equation}
where $R$ is the universal gas constant and $N_A$ is Avogadro's constant.
\newline

\noindent (iii) Orbital characteristics
\newline
\indent Earth's position and orientation relative to the Sun are not static, but vary sufficiently over time to impact the planet's climate. In particular, eccentricity, precession and obliquity changes (the Milankovitch Cycles) are known to have a major impact on climate over geological timespans (Spiegel \textit{et al.}, 2010). Precession (the rotation of the Earth's axial tilt) occurs on a cycle of approximately 19,000 years, obliquity varies between 22.1$^{\circ}$ and 24.5$^{\circ}$ approximately every 41,000 years and eccentricity cycles between 0 and 0.06 approximately every 100,000 years (Paillard, 2010; V\'{a}squez \textit{et al.}, 2010).
\newline
\indent The influence of obliquity variations on the insolation received at a given latitude can be found from
\begin{equation}
S=\frac{S_0}{\pi}\left(\frac{d_E}{d}\right)^2\left[h_0\mbox{sin}(\lambda)\mbox{sin}(\delta)+\mbox{cos}(\lambda)\mbox{cos}(\delta)\mbox{sin}(h_0)\right]
\end{equation}
where $d$ is the Earth-Sun distance at a given point in a the planet\rq{}s orbit, $h_0$ is the solar hour angle at which insolation at a particular latitude becomes positive, $\lambda$ is latitude and $\delta$ is the solar declination, which depends on obliquity ($\phi$) such that
\begin{equation}
\mbox{sin}(\delta)=\mbox{sin}(\phi)\mbox{sin}(\theta-\omega)
\end{equation}
where $\theta$ is the orbital longitude of the planet and $\omega$ is the longitude of perihelion. The eccentricity ($e$) of the planet\rq{}s orbit influences the planet-star distance ($d$) at any point in the orbit following the relation
\begin{equation}
d=\frac{d_E(1-e^2)}{(1+e\mbox{cos}(\nu))}
\end{equation}
where $\nu$ is the true anomaly of the orbit, i.e. the angle formed by the line joining a planet and star and the central axis of the ellipse formed by the planet's orbit.

\indent While they provide a guide as to the likely behaviour of the planet's orbital characteristics from the present onwards, the Milankovitch cycles are unlikely to remain constant for the entire lifetime of the planet. The presence of a large moon helps to stabilise Earth's obliquity, preventing large obliquity swings like those experienced by Mars for example (Laskar \textit{et al.}, 1993b; Laskar \textit{et al.}, 2004). While the presence of a large moon may not be essential for a stable obliquity range for Earth-like planets (Lissauer \textit{et al.}, 2012), the recession of an already present large moon could potentially induce larger obliquity swings, effecting latitudinal temperature values. Earth's moon is receding due to tidal interactions with Earth at a current rate of approximately 4 cm/yr (N\'{e}ron de Surgy \& Laskar, 1997), placing it 40,000 km further away within 1 Gyr (if this recession rate remains constant). This gives it an orbital distance of 67 R$_{\oplus}$ (R$_{\oplus} =$ Earth radii), exceeding the critical point described by Tomasella \textit{et al.} (1996) for stabilising planetary obliquity. Exceeding this distance places the Earth into a chaotic obliquity regime permitting large obliquity swings between 30-60$^{\circ}$ (Tomasella \textit{et al.}, 1996) and possibly allowing even larger obliquities of up to 90$^{\circ}$ (Laskar \textit{et al.}, 1993). This would effect global temperature distribution, causing shifts in the latitude zones that receive maximum and minimum insolation. As the Moon recedes, Earth\rq{}s rotation rate will slow. This reduces the equator-to-pole temperature gradient by reducing the magnitude of the mid-latitude eddies responsible for heat transport (Feulner, 2012), again influencing the global temperature distribution.
\newline
\indent Similarly, eccentricity will not remain stable over geological timescales. Laskar \& Gastineau (2009) suggest that a resonance between Mercury and Jupiter may eventually increase Mercury's eccentricity, which would destabilise the inner planets in approximately 3 Gyr - a scenario that could lead to Earth's eccentricity reaching a value as high as 0.3. However, this change is likely to come too late to influence Earth\rq{}s final life.
\newline
\indent Therefore, the evolution of orbital characteristics will influence climate evolution and hence, the evolution of planetary habitability. Thus, these factors need to be included in a habitability evolution model.
\newline

\noindent (iv) Hydrogen Escape and the Runaway Greenhouse Effect
\newline
\indent Surface temperatures and incoming radiation can drive positive feedback cycles that eventually send a planet's climate out of radiative equilibrium, potentially causing abrupt environmental changes. In particular, runaway and moist greenhouse effects can put an end to a planet's habitable life.
\newline
\indent As the luminosity of a star increases, the radiation intercepted by an orbiting planet increases. This increases evaporation rates of surface water, raising the atmospheric water vapour content. Water vapour being a greenhouse gas, this results in further temperature increase, beginning a positive feedback loop (McGuffie \& Henderson-Sellers, 2005; Goldblatt \& Watson, 2012). A moist greenhouse  state describes the state in which water vapour dominates the troposphere while the water vapour content of the stratosphere starts to increase. A runaway greenhouse scenario occurs when water vapour becomes a dominant component of the atmosphere. At this point, the moist adiabatic lapse rate (the net rate of temperature increase with height for an atmosphere saturated with water vapour) tends towards the saturation vapour pressure curve for water (i.e. the pressure at which water vapour becomes saturated for a given temperature) resulting in a fixed temperature-pressure structure for the atmosphere (Barnes \textit{et al.}, 2012; Goldblatt \& Watson, 2012).
\newline
\indent A fixed temperature-pressure structure means that the energy radiated back into space by the planet is also fixed. If the incoming radiation exceeds this limit, there is a net gain of energy by the planet and, provided that water vapour is not lost from the atmosphere, runaway heating occurs, eventually resulting in the evaporation of surface oceans. The oceans are finally lost to space through hydrogen escape. Ordinarily, on Earth, little water vapour reaches the stratosphere as the current temperature-pressure structure leads to water condensing at lower altitudes. However, during a runaway greenhouse, the entire atmosphere becomes water saturated, causing photo-dissociation of water molecules in the upper atmosphere to contribute significantly to the hydrogen escape. 
\newline
\indent In this case, hydrogen loss to space would be approximately $6.6\times 10^{11}$ atoms per cm$^2$ per second, limited only by the solar extreme UV heating rate (Caldeira \& Kasting, 1992), the intensity of which should be reasonably similar to the present flux for an ageing sun-like star (see, for example Ribas \textit{et al.}, 2005). Hence, assuming Hydrogen loss is uniform across the surface area of the outer atmosphere (assuming the surface area of a sphere of radius 1R$_{\oplus}$), $3.06\times 10^{30}$ hydrogen atoms would be lost per second. There are approximately $4.4\times 10^{46}$ water molecules in the ocean ($8.8 \times 10^{46}$ H atoms), therefore all water would be lost to space in approximately 1 Gyr.
\newline

\begin{figure}
                    \centering
		\includegraphics[scale=1.0]{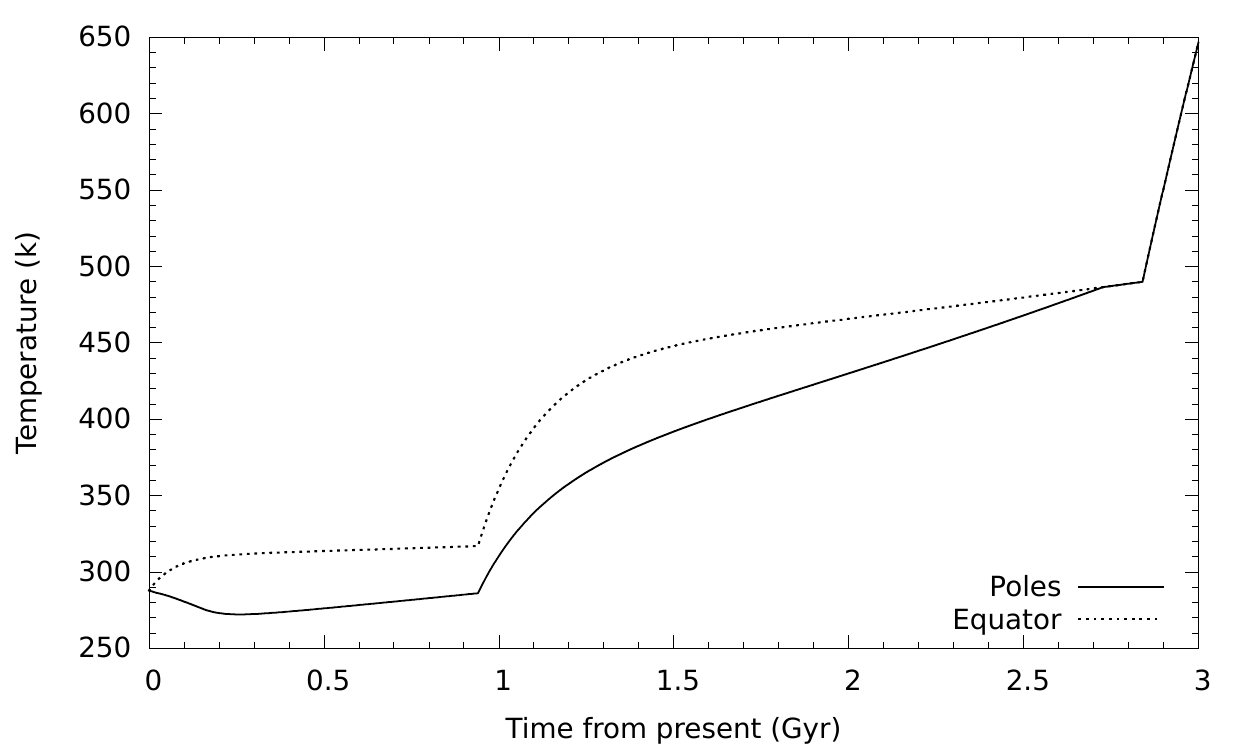}
           \caption{\textit{Change in global mean temperatures over time with increasing solar luminosity. The dashed line represents equatorial temperature and the solid line represents polar temperature. After about 1 Gyr a moist greenhouse begins when temperatures reach 330 K allowing the water vapour content of the stratosphere to increase rapidly. When temperatures reach approximately 420 K life would likely no longer be able to survive. A runaway greenhouse regime begins after approximately 2.8 Gyr. Initially, the poles warm noticeably less rapidly than the equator; however, as the planet heats up, the equator-to-pole temperature gradient decreases due to an increased latitudinal heat diffusion coefficient, caused by the increase in atmospheric pressure.}\label{Fig:fig5}}
\end{figure}

\noindent\textbf{RESULTS \& DISCUSSION}
\newline
\indent\textit{Figure \ref{Fig:fig5}} shows the general mean surface temperature trends predicted by the model for the next 2.5 Gyr. The steep increase in temperature at around 1 Gyr from the present represents the onset of rapid ocean evaporation. Assuming an upper temperature bound for life of 420 K (allowing some increase over the currently known upper temperature tolerance of thermophiles along the lines of Ward \& Brownlee (2002)) and assuming no changes to obliquity or eccentricity cycles, life could persist 0.7 Gyr longer at surface levels at the poles than at the equator.
\newline
\indent Rapid ocean loss as a result of a moist greenhouse effect would likely represent the end-point of a planet's habitable lifetime. Assuming ocean loss was not uniform across the globe due to regional temperature variations, there could potentially be pockets of liquid water that remain for a brief time before total loss of liquid surface water occurs. A source of liquid water is a prerequisite for life as we know it; hence, these last pools of water would represent the final habitable regions on a dying planet. In this section, potential locations for these last habitable regions are discussed.
\newline

\noindent (i) Ocean floor trenches
\newline
\indent The present mean ocean depth is approximately 4 km; however, deep ocean trenches, depressions in the sea floor caused by the subduction of one tectonic plate under another at divergent plate boundaries can extend the ocean depth to 6 km on average and up to 11 km for the deepest known trench, the Mariana Trench in the Western Pacific Ocean.
\newline
\indent One could imagine that as surface water is lost these would provide sheltered regions, cooler than the ambient temperatures outside the trench, much like the shadowing effect of craters on the Moon can reduce the temperatures within them by as much as 295 K compared to surface temperatures (Margot \textit{et al.}, 1999). However, the presence of an atmosphere on Earth complicates the situation. Movement of air into a trench from above would cause that air to be compressed the lower it goes. This compression increases the air temperature, which could lead to trenches actually being some of the warmest regions on a runaway greenhouse planet and thus, making them unlikely candidates for the location of some of the last liquid surface water on the planet.
\newline
\indent Additionally, it is possible that trenches may not even be present 1 Gyr from now. Plate tectonics is driven by the transfer of heat from the Earth's core to the surface. As described in more detail in Meadows (2007), heat is transferred from the core through the mantle via convection (and sometimes conduction). As these plumes of heat reach the crust they give up their heat and are pushed aside by new hot uprisings. This sideways motion in the upper mantle is the driving force for plate movement. However, the Earth is cooling down over time, resulting in a gradually solidifying outer liquid core. This slows down convection within the mantle, leaving less power available to move plates. Eventually (approximately 3 Gyr from now), convection will stop altogether ending plate motions.
\newline
\indent This slowing of tectonic processes may mean that there may not be any deep ocean trench features on the far future Earth. Additionally, regardless of whether convection is still occurring in the mantle, once the oceans boil away, tectonic plate movements will stop. The subduction of one plate under another is aided by the presence of liquid surface water, which acts as a lubricant. Without it, plate movements cease due to increased friction (Meadows, 2007).
\newline

\noindent (ii) High altitude pools
\newline
\begin{figure}
                    \centering
		\includegraphics[scale=1.0]{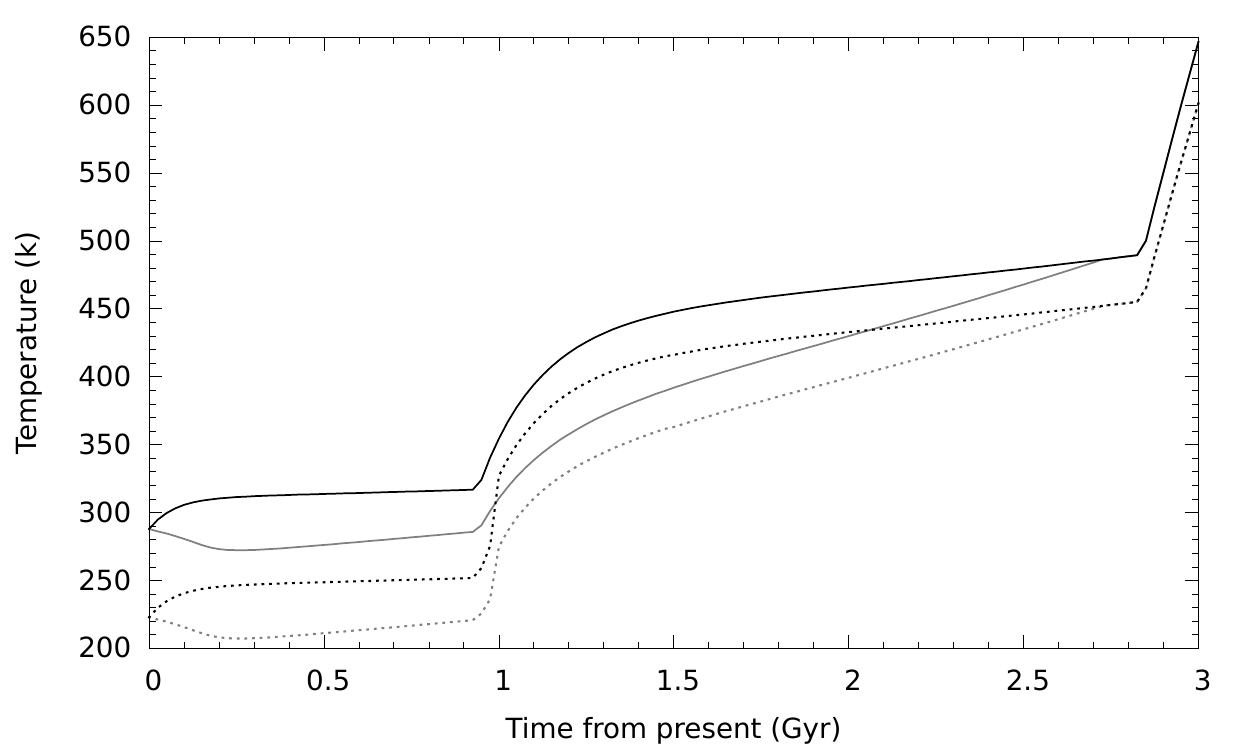}
           \caption{\textit{Mean temperature evolution at the equator (black) and the poles (grey) with increasing altitude and estimated tropospheric lapse rate from surface level (solid line) to an altitude of 10 km (dashed line).}\label{Fig:fig6}}
\end{figure}
\indent Another potential refuge could be high altitude lakes. Currently, temperatures in the troposphere follow a mean linear decrease of $\sim$6.5 K per km with increasing altitude (an average of the lapse rates calculated for dry and moist air) due to the fact that solar radiation is absorbed by the planet\rq{}s surface, which re-radiates it, heating the lower atmosphere. Assuming this negative trend still holds on the far future Earth, life could be expected to migrate upwards to more comfortable, liquid water permissible temperatures as surface temperatures rise.
\newline
\indent An estimate of the likely value of the (adiabatic) lapse rate $\Gamma_w$ during the planet's moist greenhouse phase can be obtained from
\begin{equation}
\Gamma_w = g\left[\frac{1+H_vr/R_{sd}T}{C_p+H_v^2r\epsilon/R_{sd}T^2}\right]
\end{equation}
where $g$ is the gravitational acceleration, $H_v$ is the heat of vaporisation for water, $R_{sd}$ is the specific gas constant for dry air, $\epsilon$ is the ratio of the specific gas constant of dry air to that for water, $T$ is the surface temperature and $r$ is the ratio of the mass of water vapour to dry air, which depends on the sautrated vapour pressure and atmopsheric pressure.
\newline
\indent At an altitude of 10 km (with no changes to obliquity or eccentricity cycles) life could persist for approximately 0.7 Gyr longer at polar latitudes than in equatorial regions (assuming an upper temperature tolerance of 420 K) - see \textit{Figure \ref{Fig:fig6}}. Polar surface life could outlast equatorial high-altitude life by approximately 0.2 Gyr.
\newline
\indent The case for high altitude pools may be complicated by the slowing, or halting of plate tectonics. This would mean that, except in regions where magma plumes continue to rise to the surface, continental uplift could slow down or stop, allowing the weathering rate to exceed the mountain-building rate, potentially resulting in less high altitude land area (Meadows, 2007). However, the lack of plate movements could allow volcanoes surrounding magma plumes (hot spot volcanoes) to grow taller than those on the present Earth as material is able to accumulate in one location for a longer period of time - the driver behind very tall martian volcanoes, such as Olympus Mons, which stands at a height of 21 km (Wood, 1984).
\newline
\indent Additionally, an airborne biosphere of microorganisms could also take advantage of the cooler temperatures at higher altitudes. Microorganisms are known to currently be present in the atmosphere (Bowers \textit{et al.}, 2009; Womack \textit{et al.}, 2010), some of which remain metabolically active and have residence times in the atmosphere that are long enough for them to reproduce (Womack \textit{et al.}, 2010).
\newline

\noindent (iii) Caves
\newline
\indent It can be argued that life on the far future Earth would retreat underground and into cave systems (the subsurface biosphere) (Irwin \& Schulze-Makuch, 2011). Many microbial communities are already known to exist independently of solar energy, for example, obtaining the means to metabolise directly from the rocks (\textit{chemolithotrophs}) and in the case of one strain of green sulphur bacteria, even using the light from geothermal radiation around a deep sea vent for photosynthesis (Beatty \textit{et al.}, 2005).

\indent Caves are generally assumed to have constant interior environments, with internal temperatures closely approximating the local mean surface temperature (Tuttle \& Stevenson, 1977; Howarth, 1983). This, combined with the general trend of increasing temperatures with depth would not bode well for organisms on  a planet with surface temperatures exceeding the boiling point of water. However, some cave systems may be better suited to sheltering life than others. In particular, caves that have their greatest volume below their entrance would act as cold reservoirs (Tuttle \& Stevenson, 1977) as colder, denser air would flow downward into the cave, but warmer lighter air would not. 
\newline
\indent These cold trap caves, (also known as ice caves due to the presence of year-round ice within them) are often formed from collapsed lava tubes (Williams \textit{et al.}, 2010) and generally have a narrow single entrance with a large chamber below, which has a large volume-to-wall surface ratio, because it is conduction through the cave walls that would restore the temperature within the cave to the mean surface temperature. 
\newline
\indent Cold, dense air enters the cave during winter. In-falling snow is compacted into layers of ice, or inflowing water freezes. When temperatures outside increase, the warmer, less dense air cannot enter the cave, leaving the colder air trapped within. This allows ice to remain within the cave throughout the year. However, the ice mass within the cave is not the ice that originally formed there. The ice is continually being melted as a result of heat conduction from the surrounding rocks that are in contact with the ice (which tend to reflect the annual mean temperature of a region) and dripping water inside the cave (Ohata \textit{et al.}, 1994; Leutscher \textit{et al.}, 2005). The melt water is then lost from the cave with the entire ice mass being replaced after between 100-1000 years (Leutscher \textit{et al.}, 2005), depending on the geometry of the particular cave in question. Hence, new 
\noindent water needs to be supplied ($\Delta w$) such that
\begin{equation}
\Delta w = \Delta w_{in} - \Delta w_{melted}
\end{equation}
remains constant for permanent ice to remain. 
\newline
\indent Therefore, for this mechanism to provide a useful refuge for life on the far-future Earth, there would need to be some form of water input into the system (perhaps through seepage from the surrounding rock or via water vapour entering from the external atmosphere at times when the outside air density  matches, or exceeds that within the cave) for liquid water to remain within such a cave.
\newline
\indent Ice caves are found not only in cold regions, but also temperate and tropical latitudes on Earth today. The question of how long such caves would be able to retain water when exposed to rapidly increasing surface temperatures remains open. Leutscher \textit{et al.} (2005) focused on ice caves in the Jura Mountains and found that the temperatures within cold trap caves were dependent on the outside winter temperatures in the region and, more particularly relevant  to this work, that the ice mass within the ice caves studied decreased as annual mean winter temperatures increased (most noticeably since the 1980s). This suggests that cold trap cave temperatures on the far-future Earth would only be as (relatively) cool as the coolest temperatures in a region in any given year, which, given the high temperatures expected, may not provide a long stay of execution for life. 
\newline
\begin{figure}
                    \centering
		\includegraphics[scale=1.3]{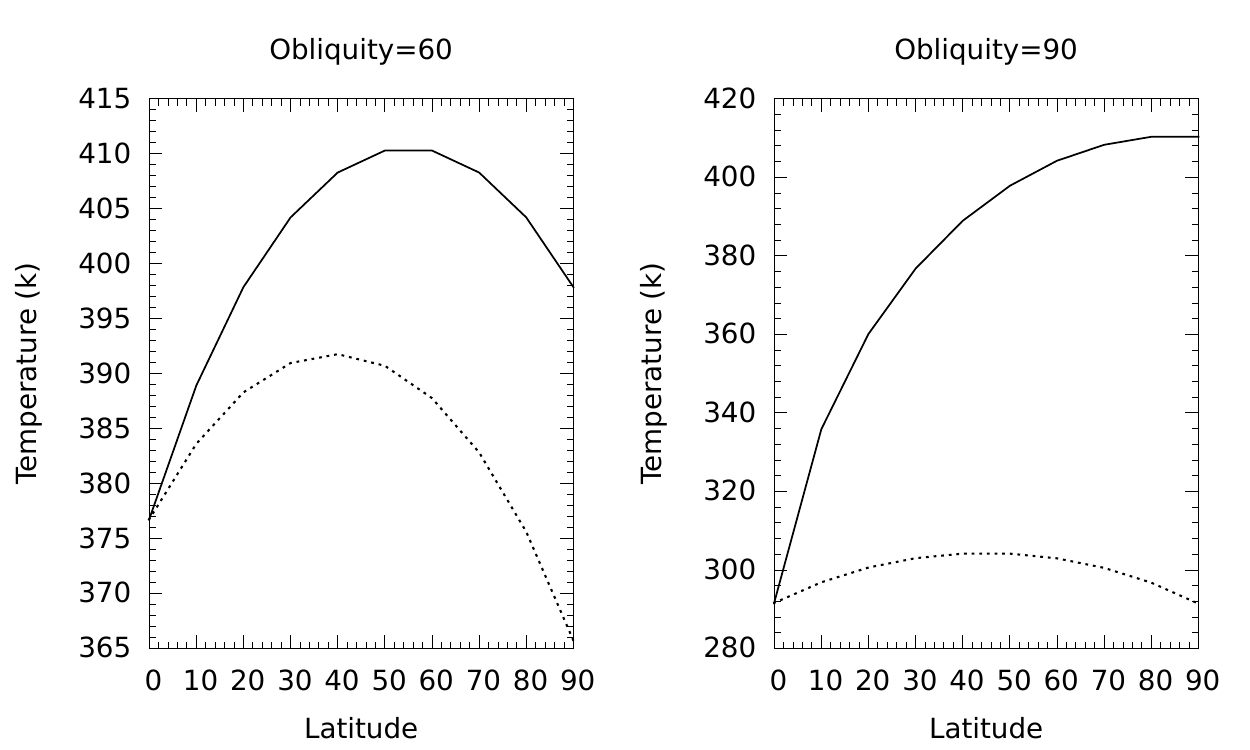}
           \caption{\textit{Seasonal temperature variations between equatorial and polar latitudes for obliquities of 60 and 90 degrees for a solar luminosity of 1.1 L$_{\odot}$. The solid lines represent the upper temperature range and the dashed lines represent the lower temperature range.}\label{Fig:fig7}}
\end{figure}
\indent However, for Earth, after the Moon recedes beyond the critical distance of 67 $R_{\oplus}$ described by Tomasella \textit{et al.} (2006) (approximately 1 Gyr from now), Earth\rq{}s obliquity enters a regime in which it can vary chaotically between much higher values than the present obliquity range. If obliquity cycling in the model is altered to vary between 30-60$^{\circ}$ (and even up to 90$^{\circ}$), the equator-to-pole temperature gradient is effectively reversed by the new, higher obliquity range, which causes the poles to receive more insolation than the equator (in agreement with Tomasella \textit{et al.} (2006)). As such high obliquity planets do not necessarily preclude habitability (Williams \& Kasting, 1997) this suggests that equatorial regions may be more accommodating to life than polar regions on the far-future Earth.
\newline
\indent These large obliquity swings raise the possibility that the cold trap cave mechanism could extend the lifetime of liquid water habitats further than expected. The more extreme temperature differences between seasons that result from increased obliquity would cause cooler winter temperatures. For example, as illustrated in \textit{Figure \ref{Fig:fig7}}, an obliquity of 60$^{\circ}$ would allow winter temperatures that are up to 40 degrees lower than summer temperatures in some regions, whereas an obliquity of 90$^{\circ}$ could potentially allow very extreme seasonal temperature variations of more than 200 degrees. This, of course, neglects latitudinal heat transport, which acts to lower the temperature range (especially as atmospheric density increases). By including simple heat transport, an estimate of the magnitude of the lower temperature values over latitude for different obliquity values was found (\textit{Figure \ref{Fig:fig8}}). A high obliquity far-future Earth with a cold-trap cave could extend the stay of execution for life as far as $+ 2.8$ Gyr from present.
\begin{figure}
                    \centering
		\includegraphics[scale=1.2]{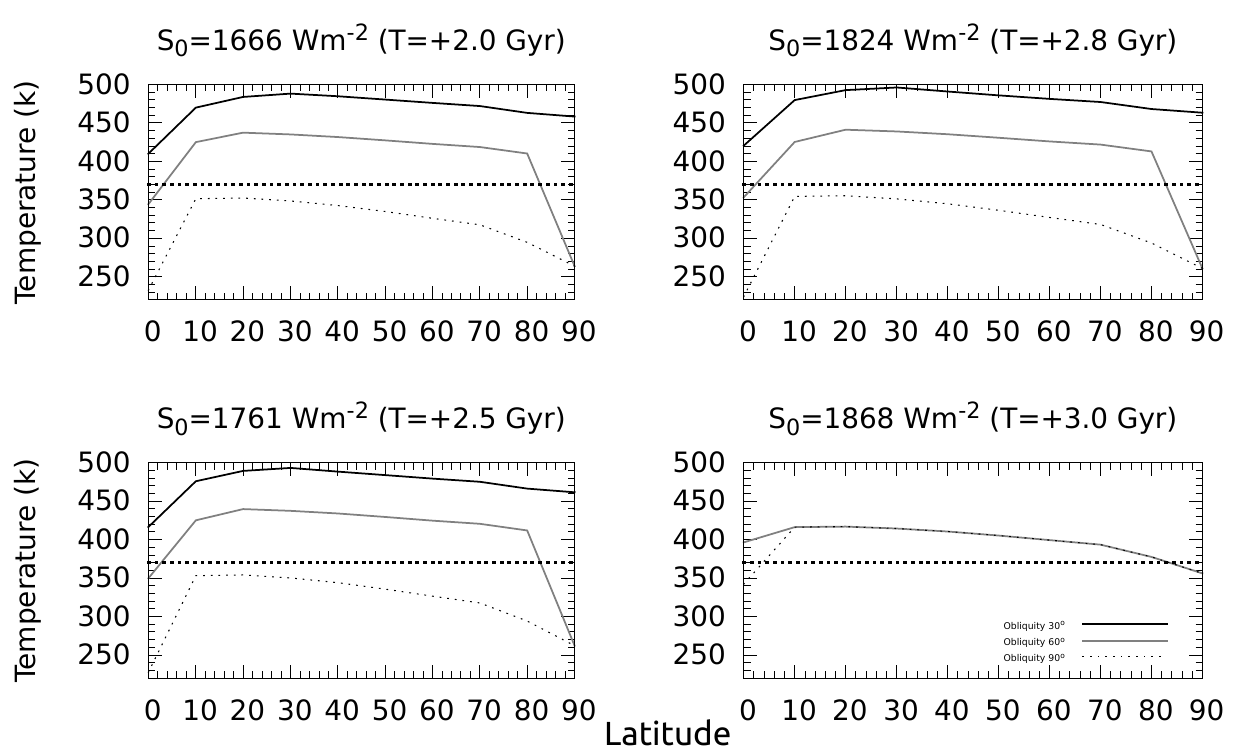}
           \caption{\textit{Change in lower temperature range over latitude for obliquities of  30, 60 and 90 degrees at 2.0, 2.5, 2.8 and 3.0 Gyr from present. These suggest that, for high obliquities, temperatures may permit liquid water at some, or all latitudes at certain points in the planet\rq{}s orbit for at least 2.8 Gyr from present, which would allow for a cold trap cave mechanism to harbour liquid water year-round in these cases. While the $+$3 Gyr example shows a slight dip below the boiling point of water for high/low latitudes for obliquities of 60/90 degrees, the onset of a runaway greenhouse would lead to much less predictable climate dynamics and water availability would be very limited.}\label{Fig:fig8}}
\end{figure}
\newline

\noindent (iv) The deep subsurface
\newline
\indent The deep-subsurface environment would likely provide more stable refuges for life. It has been claimed that the deep biosphere contains the vast majority of Earth\rq{}s microbial biomass, with 50-80$\%$ of this estimated to occur in the deep marine subsurface (Orcutt \textit{et al.}, 2011). It is officially defined as depths of 50 m or more below the continental surface or the ocean floor (Bell, 2012) although, in practice, deep biosphere research can encompass any depths below 1 m (Bell, 2012; Edwards \textit{et al.}, 2012). The lower depth limit is currently unknown (Kieft \textit{et al.}, 2005; Reith, 2011). The limiting factor is likely to be temperature. It is known that temperature increases with depth, with the mean geothermal temperature gradient for Earth\rq{}s crust being $\sim$30 degrees per km depth (Bohlen, 1987). This would place the lower limit at the depth where temperatures reach the upper-limit for life (122$^{\circ}$C) - approximately 3.5 km for the present Earth. 

However, the geothermal gradient varies depending on rock type (Pedersen, 2000), increasing or decreasing the maximum habitable depth in a given region. The deepest microbial life found to date was found in igneous rock aquifers at depths of 5.3 km. Therefore, although assuming a mean geothermal gradient of 30 degrees would lead to predicted temperatures that preclude habitability once surface temperatures reach the maximum for life, the variable nature of the crustal temperature gradient could allow the deep subsurface biosphere to persist, perhaps for longer than any other refuge.

In the deep subsurface biosphere life has been characterised that operates independently of surface photosynthesis (Kieft \textit{et al.}, 2005; Lin \textit{et al.}, 2005; Lin \textit{et al., 2006}; Chivian \textit{et al.}, 2008). From a remote detection perspective it may be more challenging to detect the presence of life in these environments; however, life existing independently of photosynthesis might produce methane, which could be detectable (Parnell \textit{et al.}, 2010).
\newline

\noindent\textbf{Future life in hot, saline environments?}
\newline
\indent Wherever these last pools of water may be found, they are likely to share some common characteristics. They are likely to be warm, isolated and (at least for ocean remnants) highly saline. This suggests that the last lifeforms would be those that can tolerate, or require high salt concentrations and temperatures (\textit{thermohalophiles}).
\newline
\indent Life on the far-future Earth will likely be found in microbial mat communities resembling some present day hydrothermal vent communities (Ward \& Brownlee, 2002). High temperature, high salinity hydrothermal vent environments are found on Earth today. For example, the shallow-water Milos submarine hydrothermal field in the central Hellenic Volcanic Arc provides such an environment. The high salinity of some waters in this region (as much as 2.5 times higher than ordinary sea water (Valsami-Jones \textit{et al.}, 2005)) means that only microbial mats can thrive in those areas immediately surrounding hydrothermal vents (Thiermann \textit{et al.}, 1997).
\newline
\indent It is not possible to say with any certainty what the likely pH of the final pools of liquid water would be. Alkaliphilic and moderately acid-tolerant forms of halophlic microorganisms are known on Earth today, spanning an optimal pH range from 4-10 (DasSarma \& DasSarma, 2012), whereas thermophiles are found across the pH range from 1-10 (Ferrara \& Reysenbach, 2007). Therefore, although the pH of these far-future environments cannot be determined, the required survival mechanisms for high temperature, high salinity environments should be able to adapt to function.
\newline
\indent Organisms living in this future epoch would also be subjected to elevated UV radiation levels due to the low abundance of atmospheric ozone. Ozone is produced from O$_2$ and when atmospheric O$_2$ decreases, as a result of the lack of O$_2$ production when temperatures exceed the upper limit for oxygenic photosynthesis, the steady state level of ozone decreases, resulting in less attenuation of biologically damaging UV radiation below 290-300 nm (Beaty \textit{et al.}, 2006)
\newline
\indent Recently characterised microbiological communities from some of the highest Atacaman volcanoes (Lynch \textit{et al.}, 2012) may reflect the style of microbial communities expected on a hot, dry Earth. These exist under a very intense solar radiation regime, with large ($\Delta T > 60^{\circ}$C) daily temperature variations and under very arid conditions. Known examples of microbes adapted to high salinities and warm temperatures are \textit{Halothermothrix orenii}, \textit{Thermohalobacter berrensis} and \textit{Bacillus aeolius}: these provide plausible models for the last life on Earth. Halobacteria  and many other salt-dwelling Archaea and Bacteria often inhabit salt crusts, which confer protection from deleterious UV radiation (Landis, 2001) and one could imagine salt-encrusted communities of \textit{thermohalophiles} inhabiting the last remaining bodies of water. The supposition that such life would be halophilic depends, of course, on the chemical composition of the water.
\newline
  \begin{table}
   \begin{center}
    \begin{tabular}{ | l | r|}
    \hline
    \textbf{Era} & \textbf{Productivity} ($\times 10^{12}$ mol C yr$^{-1}$) \\ \hline\hline
    Early Earth & 180-560 (Canfield, 2005)  \\ \hline
    Modern Earth & 8740 (Field \textit{et al.}, 1998)  \\ \hline
    \end{tabular}
   \caption{Net global biological productivity values for modern and early (pre-photosynthetic) Earth.}
  \label{tab:Tab1}
  \end{center}
  \end{table}

\noindent\textbf{Future biosignatures?}
\newline
\indent This leads to the question of what the dominant by-products of such an extremophile biosphere would be and whether this would allow for any remotely detectable atmospheric biosignatures. The microorganisms from the Atacaman volcanoes appear to use the oxidation of carbon monoxide to obtain energy (Lynch \textit{et al.}, 2012). While the slowing of plate tectonics may decrease the amount of available CO, hotspot volcanoes (those caused by upwellings from the deep mantle rather than at plate boundaries) could still exist. Mars (which is not tectonically active) exhibited this kind of volcanism up to as early as 2 Myr ago (Neukum \textit{et al.}, 2004), while hotspot volcanism may still be actively ongoing on Venus (Smrekar \textit{et al.}, 2010). The thermohalophilic organisms listed above are all anaerobic organisms which produce hydrogen, carbon dioxide, ethanol and acetate as by-products of their metabolic processes. Acetates are not promising as remotely detectable biosignatures, only managing to escape into the atmosphere (at a very slow rate) if they are formed within very acidic (pH $<$ 4.7) water (Seager \textit{et al.}, 2012). 
\newline
\indent On a planet with relatively low carbon dioxide levels compared to the present Earth, it may be possible that the biological production of CO$_2$ would produce a disequilibrium of the gas in the atmosphere that would indicate the presence of life, especially if it can easily be deduced that the planet is likely to be near the end of its habitable lifetime. Similarly, if enough hydrogen is produced biologically, there could be a noticeable excess of atmospheric hydrogen, if biological hydrogen production is not comparatively negligible to that produced abiotically.
\newline
  \begin{table}
   \begin{center}
    \begin{tabular}{ | l  l | r| c| }
    \hline
    \textbf{Refuge} & ~ & \textbf{Productivity} ($\times 10^{12}$ mol C yr$^{-1}$) & \textbf{Source} \\ \hline\hline
    Pools/lakes & Eutrophic & (2.2 - 15)$\times 10^{-12}$ [per m$^3$ water]  & Nixdorf \textit{et al.} (2003) \\
    ~ & Acidic &  4.7$\times 10^{-12}$ [per m$^3$ water] &  \\
    ~ & Warm & (3.6 - 4.5)$\times 10^{-12}$ [per m$^3$ water]  &Kimura \textit{et al.} (2010) \\ \hline
   Caves & Chemoautotrophic & 2.2 $\times 10^{-18}$ [per mg dry weight] & Engel \textit{et al.} (2001)  \\
    ~ & Heterotrophic & 0.1  $\times 10^{-18}$ [per mg dry weight] & \\ \hline
    Deep subsurface & Hydrothermal vent &  5.1 $\times 10^{-6}$ [per vent] & McCollom (1999) \\ 
    ~ & Basaltic crust & 0.08 &  Bach \& Edwards (2003)\\ \hline
    \end{tabular}
   \caption{Biological productivity values for environments similar to the proposed refugia on the far-future Earth.}
  \label{tab:Tab2}
  \end{center}
  \end{table}
\indent Global biological productivity on the early (pre-photosynthetic) Earth was much lower than that present-day value, as illustrated by the values in Table \ref{tab:Tab1}. Similarly, it is likely that productivity on the far-future Earth will be much lower than at present. An estimate for the likely productivity of the planet towards the end of its habitable lifetime is obtained by looking at the productivity values for environments similar to the proposed refugia for life (summarised in Table \ref{tab:Tab2}).
\newline
\indent As a best-case scenario, assuming a global distribution of geothermal pools equal in size to the largest on the present-day Earth (Frying Pan Lake, New Zealand - 200,000 m$^3$ (Scott, 1994)), a distribution of ice caves similar to that in the Jura Mountains (40 caves (Leutscher \textit{et al.}, 2005)) for 10 mountain ranges, a similar number of active hydrothermal vent fields (there are estimated to be approximately 1000 currently active vent fields (Baker \& German, 2004)) and a microbially-inhabited basaltic crust, a guideline figure for global net productivity of 0.23 ($\times 10^{12}$ mol C yr$^{-1}$) is obtained for the far-future Earth; two orders of magnitude lower than the pre-photosynthetic Earth. This suggests that the biosignatures produced by the far-future biosphere would be less intense than those on a world with higher global net biological productivity.
\newline

\noindent\textbf{CONCLUSIONS}
\newline
\indent As the luminosity of the Sun increases during its main sequence lifetime, surface temperatures on Earth will increase, leading to complete evaporation of the Earth\rq{}s oceans and an end to the Earth\rq{}s habitable lifetime. As the planet nears the end of its habitable lifetime, it will likely only support microbial life; more specifically, life adapted to survive in harsh high-temperature, high-salinity environments. A maximum lifetime for life on Earth of 2.8 Gyr from present was found, given the presence of sheltered, high-altitude or high-latitude environments. These niches should accommodate life for about 1 Gyr beyond other surface environments.
\newline
\indent A model was developed to investigate the effects such life will have on its environment. Future work will extend this to investigate the likely biosignatures produced by life in these localised specific environments, both on Earth and on any habitable exo-planet in the late stages of its development to answer the questions of whether a vastly reduced biosphere could produce remotely detectable biosignatures and, by extension, at what point an Earth-like planet would stop exhibiting remotely detectable biosignatures.
\newline

\noindent\textbf{Acknowledgements}
\newline
The authors wish to thank Jelte Harnmeijer, Christiane Helling and Anja Bauermeister for helpful discussions and Shawn Domagal-Goldman and an anonymous reviewer for comments that signiﬁcantly improved the manuscript. Support from the Science and Technology Facilities Council (STFC) Aurora scheme is gratefully acknowledged. The University of Dundee is a registered Scottish Charity, No. SC015096.
\newline

\noindent\textbf{References}
\small

Allan R.P. (2011) The Role of Water Vapour in Earths Energy Flows.  \textit{Surv. Geophys.} In press, doi: 10.1007/s10712011-9157-8. Online preprint from http://www.springerlink.com/content/f35156772k180362/.

Ar\'{a}ujo M.B., Thuiller W. and Pearson R.G. (2006) Climate warming and the decline of amphibians and reptiles in Europe. \textit{J. Biogeogr.} 33, 1712-1728.

Bach W. and Edwards K.J. (2003) Iron and sulﬁde oxidation within the basaltic ocean crust: Implications for chemolithoautotrophic microbial biomass production. \textit{Geochimica et Cosmochimica Acta} 67, 3871-3887.

Baker E.T. and German C.R. (2004) On the global distribution of hydrothermal vent fields. \textit{Geophysical Monograph Series} 148, 245-266.

Bar-Even A., Noor E., Lewis N.E. and Milo R. (2010) Design and analysis of synthetic carbon fixation pathways. \textit{PNAS} 107, 8889-8894.

Bar-Even A., Noor E. and Mil R. (2012) A survey of carbon fixation pathways through a quantitative lens.  \textit{J. Exp. Bot.} 63, 2325-2342.

Barnes R., Mullins K., Goldblatt C., Meadows V.S., Kasting J.F. and Heller R (2012) Tidal Venuses: Triggering a Climate Catastrophe via Tidal Heating.  \textit{Arxiv Preprint:} arXiv:1203.5104v1.

Beatty J.T., Oevrmann J., Lince M.T., Manske A.K., Lang A.S., Blankenship R.E., Van Dover C.L., Martinson T.A. and Plumley F.G. (2005) An obligately photosynthetic bacterial anaerobe from a deep-sea hydrothermal vent.  \textit{PNAS} 102, 9306-9310.

Beaty D. et al. (2006) Unpublished white paper, 76 p, posted June 2006 by the Mars Exploration Program Analysis Group (MEPAG) at http://mepag.jpl.nasa.gov/reports/index.html.

Bell E.M. (2012) Life at Extremes: Environments, Organisms and Startegies for Survival. \textit{CABI}, Oxfordshire, UK.

Birmingham B.C. and Colman B. (1979) Measurement of Carbon Dioxide Compensation Points of Freshwater Algae. \textit{Plant Physiol.} 64, 892-895.

Boer G.J., Hamilton K. and Zhu W. (2004) Climate sensitivity and climate change under strong forcing. \textit{Climate Dynamics} 24, 685-700.

Bohlen S.R. (1987) Pressure-Temperature-Time Paths and a Tectonic Model for the Evolution of Granulites. \textit{J. Geol.} 95, 617-632.

Bonfils X., Delfosse X., Udry S., Forveille T., Mayor M., Perrier C., Bouchy F., Gillon M., Lovis C., Pepe F., Queloz D., Santos N.C., S\'{e}gransan D. and Bertaux J.-L. (2011) The HARPS search for southern extra-solar planets XXXI. The M-dwarf sample.  \textit{Arxiv Preprint:} arXiv:1111.5019.

Bounama C., Franck S. and von Bloh W. (2001) The fate of Earth’s ocean.  \textit{Hydrology Earth Syst. Sci}. 5, 569-575.

Bowers R.M., Lauber C.L., Wiedinmyer C., Hamady M., Hallar A.G., Fall R., Knight R. and Fierer N. (2009) Characterization of Airborne Microbial Communities at a High-Elevation Site and Their Potential To Act as Atmospheric Ice Nuclei. \textit{App. Env. Microbiol.} 75, 5121-5130.

Butterfield N.J. (2000) Bangiomorpha pubescens n. gen., n. sp.: implications for the evolution of sex, multicellularity, and the Mesoproterozoic/Neoproterozoic radiation of eukaryotes.  \textit{Paleobiology} 26, 386-404.

Caldeira C . and Kasting J.F. (1992) The life span of the biosphere revisited.  \textit{Nature} 360, 721-723.

Canfield D.E. (2005) The Early History of Atmospheric Oxygen: Homage to Robert M. Garrels. \textit{Annu. Rev. Earth Planet. Sci.} 33, 1-36.

Catanzarite J. and Shao M. (2011) The Occurrence Rate of Earth-Analog Planets Orbiting Sun-like Stars.  \textit{Astrophys. J.} 738, 151-161.

Chivian D, Brodie E.L., Alm E.J., Culley D.E., Dehal P.S., DeSantis T.Z., Gihring T.M., Lapidus A., Lin L-H., Lowry S.R., Moser D.P., Richardson P.M., Southam G., Wanger G., Pratt L.M., Andersen G.L., Hazen T.C., Brockman F.J., Arkin A.P. and Onstott T.C. (2008) Environmental Genomics Reveals a Single-Species Ecosystem Deep Within Earth. \textit{Science} 322, 275-278.

Clarke A. and Rothery P. (2008) Scaling of body temperature in mammals and birds. \textit{Functional Ecology} 22, 58-67.

Cockell C.S. (2003) Impossible Extinction: Natural Catastrophes and the Supremacy of the Microbial World. Cambridge University Press, Camridge, UK.

Dartnell L. (2011) Biological constraints on habitability. \textit{A\&G} 52, 1.25-1.28.

DasSarma S. and DasSarma P. (2001) Halophiles. In  \textit{Encyclopedia of Life Sciences}, John Wiley \& Sons, Ltd: Chichester.

Dillon M.E., Wang G. and Huey R.B. (2010) Global metabolic impacts of recent climate warming. \textit{Nature} 467, 704-707.

Edwards K.J., Becker K. and Colwell F. (2012) The Deep, Dark Energy Biosphere: Intraterrestrial Life on Earth. \textit{Annu. Rev. Earth Planet. Sci.} 40, 551-568.

Engel A.S., Porter M.L., Kinkle B.K. and Kane T.C. (2001) Ecological Assessment and Geological Signi cance of Microbial Communities from Cesspool Cave, Virginia. \textit{Geomicrobiology Journal} 18, 259-274.

Falkowski P.G., Katz M.E., Milligan A.J., Fennel K., Cramer B.S., Aubrey M.P., Berner R.A., Novacek
M.J. and Zapol W.M. (2005) The Rise of Oxygen over the Past 205 Million Years and the Evolution of Large
Placental Mammals. \textit{Science} 309, 2202-2204.

Ferrera I. and Reysenbach A-L. (2007) Thermophiles. In Encyclopedia of Life Sciences, John Wiley \& Sons, Ltd: Chichester.

Feulner G. (2012) The faint young Sun problem.  \textit{Arxiv preprint:} arXiv:1204.4449v1.

Field C.B., Behrenfeld M.J., Randerson J.T. and Falkowski P. (1998) Primary Production of the Biosphere: Integrating Terrestrial and Oceanic Components. \textit{Science} 281, 237-240.

Goldblatt C. and Watson J.A. (2012) The Runaway Greenhouse: implications for future climate change, geoengineering and planetary atmospheres.  \textit{ArXiv Preprint:} arXiv:1201.1593v1 .

Gough D.O. (1981) Solar Interior Structure and Luminosity Variations.  \textit{Solar Phys.} 74, 21-34.

Howarth F.G. (1983) Ecology of Cave Arthropods. \textit{Ann. Rev. Etomol.} 28, 365-289.

Irwin L.N. and Schulze-Makuch D. (2011) Cosmic Biology: How Life Could Evolve on Other Worlds. Edited by John Mason. Springer Science+Business Media, New York, NY, USA.

Kaltenegger L., Traub W.A. and Jucks K.W. (2007) Evolution of an Earth-like Planet.  \textit{Astrophys. J.} 658, 598-616.

Kasting J.F. (1988) Runaway and moist green121 C in kelvinhouse atmospheres and the evolution of earth and Venus.  \textit{Icarus} 74:472-494.

Kasting J.F. and Grinspoon D.H. (1991) The Faint Young Sun Problem. In:  \textit{The Sun in Time}, University of Arizona Press, Tuscon, AZ, pp. 447-462.

Kemp T.S. (2006) The origin of mammalian endothermy: a paradigm for the evolution of complex biological structure. \textit{Zoological Journal of the Linnean Society} 147, 473-488.

Kieft T.L., McCuddy S.M., Onstott T.C., Davidson M., Lin L-H., Mislowack B., Pratt L., Boice E., Lollar B.S., Lippmann-Pipke J., Pfiffner S.M., Phelps T.J., Gihring T, Moser D. and van Heerden A. (2005) Geochemically Generated, Energy-Rich Substrates and Indigenous Microorganisms in Deep, Ancient Groundwater. \textit{Geomicrobiol. J.} 22, 325-335.

Kimura H., Mori K., Nashimoto H., Hattori S., Yamada K., Koba K., Yoshida N. and Kato K. (2010) Biomass production and energy source of thermophiles in a Japanese alkaline geothermal pool. \textit{Microbes Environ.} 12, 480-489.

Landis G.A. (2001) Martian Water: Are There Extant Halobacteria on Mars?.  \textit{Astrobiology} 1, 161-164.

Laskar J., Joutel F. and Robutel P. (1993b) Stabilization of the Earth’s obliquity by the Moon.  \textit{Nature} 361, 615-617.

Laskar, J., Correia, A.C.M., Gastineau, M., Joutel, F., Levrard, B. and Robutel, P. (2004) Long term evolution and chaotic diffusion of the insolation quantities of Mars.  \textit{Icarus} 170, 343-364.

Laskar J. and Gastineau M. (2009) Existence of collisional trajectories of Mercury, Mars and Venus with the Earth.  \textit{Nature} 459, 817-819.

Leutscher M., Jeannin P-Y and Haeberli W. (2005) Ice Caves as an indicator of winter climate evolution: a case study from the Jura Mountains.  \textit{The Holocene} 15, 982-993.

Levenson B.P. (2011) Planet temperatures with surface cooling parameterized. \textit{Adv. Space. Res.} 47, 2044-2048.

Lin L-H., Slater G.F., Lollar B.S., Lacrampe-Couloume G. and Onstott T.C. (2005) The yield and isotopic composition of radiolytic H$_2$, a potential energy source for the deep subsurface biosphere. \textit{Geochimica et Cosmochimica Acta} 69, 893-903.

Lin L-H., Wang P-L., Rumble D., Lippmann-Pipke J., Boice E., Pratt L.M., Lollar B.S., Brodie E.L., Hazen T.C., Andersen G.L., DeSantis T.Z., Moser D.P., Kershaw D. and Onstot T.C. (2006) Long-Term Sustainability of a High-Energy, Low-Diversity Crustal Biome.  \textit{Science} 314, 479-482.

Lineweaver C.H. (2001) An Estimate of the Age Distribution of Terrestrial Planets in the Universe: Quantifying Metallicity as a Selection Effect. \textit{Icarus} 151, 307-313.

Lissauer J.J., Barnes J.W. and Chambers J.E. (2012) Obliquity variations of a moonless Earth.  \textit{Icarus} 217, 77-87.

Lorenz R.D., Lunine J.I., Withers P.G. and McKay C.P. (2001) Titan, Mars and Earth : Entropy Production by Latitudinal Heat Transport.  \textit{Geophys. Res. Lett.} 28, 415-418.

Lynch R.C., King A.J., Far\'{i}as M.E., Sowell P., Vitry C. and Schmidt S.K. (2012) The potential for microbial life in the highest elevation ($>$6000 m.a.s.l.) mineral soils of the Atacama region.  \textit{J. Geophys. Res.} doi:10.1029/2012JG001961, in press.

Maberly S.C. (1990) Exogenous Sources of Inorganic Carbon for Photosynthesis by Marine Macroalgae. \textit{J. Physcol} 26, 439-449.

Maberly S.C. (1996) Diel, episodic and seasonal changes in pH and concentrations of  inorganic carbon in a productive lake. \textit{Freshwater Biology} 35, 579-598.

Margot J.L., Campbell D.B., Jurgens R.F. and Slade M.A. (1999) Topography of the Lunar Poles from Radar Interferometry: A Survey of Cold Trap Locations.  \textit{Science} 284, 1658-1660.

McLean, D. M. (1991) A climate change mammalian population collapse mechanism, in Kainlauri, E., Johansson, A., Kurki-Suonio, I., and Geshwiler, M., eds., Energy and Environment: Atlanta, Georgia, ASHRAE, p. 93-100.

McGuffie K. and Henderson-Sellers A. (2005) A Climate Modelling Primer, 3rd Edition. John Wiley \& Sons Ltd., West Sussex, England.

Meadows A.J. (2007) The Future of the Universe. Springer-Verlag London Limited 2007.

Mesbah N.M. and Wiegel J. (2012)  Life Under Multiple Extreme Conditions: Diversity and Physiology of the Halophilic Alkalithermophile. \textit{Appl. Environ. Microbiol.} 78, 4074-4082.

McCollom T.M. (1999)  Methanogenesis as a potential source of chemical energy for primary biomass production by autotrophic organisms in hydrothermal systems on Europa. \textit{J. Geophys. Res.} 104, 30729-30742.

Mojzsis  S.J., Arrhenius G., McKeegan K.D., Harrison T.M., Nutman A.P. and Friend C.R.L. (1996) Evidence for life on earth before 3,800 million years ago. \textit{Nature} 384, 55-59.

Myhre G., Highwood E.J., Shine K.P. and Stordal F. (1998) New estimates of radiative forcing due to well mixed greenhouse gases.  \textit{Geophys. Res. Lett.} 25, 2715-2718.

N\'{e}ron de Surgy O. and Laskar J. (1997) On the long term evolution of the spin of the Earth.  \textit{Astron. \& Astrophys}. 318, 975-989.

Neukum G., Jaumann R., Hoffmann H., Hauber E., Head J.W., Basilevsky A.T., Ivanov B.A., Werner S.C., van Gasselt S., Murray J.B., McCord T. and The HRSC Co-Investigator Team (2004) Recent and episodic volcanic and glacial activity on Mars revealed by the High Resolution Stereo Camera. \textit{Nature} 432, 971-979.

Nixdorf B., Krumbeck H., Jander J. and Beulker C. (2003) Comparison of bacterial and phytoplankton productivity in extremely acidic mining lakes and eutrophic hard water lakes. \textit{Acta Oecologica} 24, S281-S288.

Ohata T., Furukawa T. and Osada K. (1994) Glacioclimatological Study of Perennial Ice in the Fuji Ice Cave, Japan. Part 2. Interannual Variation and Relation to Climate.  \textit{Arctic Alpine Res.} 26, 238-244.

Orcutt B.N., Sylvan J.B., Knab N.J. and Edwards K.J. (2011) Microbial Ecology of the Dark Ocean above, at and below the Seafloor. \textit{Microbiol. Mol. Biol. Rev.} 75, 361-422.

Oren A. (2009) Microbial Diversity. In  \textit{Encyclopedia of Life Sciences (ELS)}. John Wiley \& Sons, Ltd., Chichester.

Paillard A.A. (2010) Climate and the orbital parameters of the Earth.  \textit{C. R. Geosc.} 342, 273-285.

Parnell J., Boyce A.J. and Blamey N.J.F. (2010) Follow the methane: the search for a deep biosphere, and the case for sampling serpentinites, on Mars.  \textit{Int. J. Astrobiol.} 9, 193-200.

Pavlov A.A., Kasting J.F., Brown L.L., Rages K.A. and Freedman F. (2000) Greenhouse Warming by CH4 in the Atmosphere of Early Earth.  \textit{Geophys. Res.} 105, 11981-11990.

Pedersen K. (2000) Exploration of deep intraterrestrial microbial life: current perspectives. \textit{FEMS Microbiol. Lett.} 185, 9-16.

Price C. and Asfur M. (2006a) Can Lightning Observations be Used as an Indicator of Upper Tropospheric Water Vapor Variability?  \textit{Bull. Amer. Meteor. Soc.} 87, 291-298.

Price C. (2008) Thunderstorms, Lightning and Climate Change. 29th International Conference on Lightning Protection, 23rd-26th June 2008, Uppsala, Sweden.

R\'{a} k\'{o}czi F. and Iv\'{a}nyi S. (1999) Water vapour and greenhouse effect.  \textit{Geofizika} 16-17, 65-72.

Raven J.A. and Larkum A.W.S. (2007) Are there ecological implications from the proposed energetic restrictions on photosynthetic oxygen evolution at high oxygen concentrations? \textit{Photosynthesis Research} 94, 31-42.

Reith F. (2011) Lif in the deep subsurface. \textit{Geology} 39, 287-288.

Ribas I., Guinan E.F., G\'{u}del M. and Audard M. (2005) Evolution of Solar Activity Over Time and Effects on Planetary Atmospheres. I. High-Energy Irradiances (1-1700 A).  \textit{Astrophys. J.} 622, 680-694.

Rosing, M. T. (1999) C-13-depleted carbon microparticles in $>$ 3700-Ma sea-ﬂoor
sedimentary rocks from west Greenland. \textit{Science} 283, 674-676.

Rosing M.T., Bird D.K., Sleep N.H. and Bjerrum C.J. (2010) No climate paradox under the faint
early Sun.  \textit{Nature} 464, 744-747.

Rothschild L.J. and Mancinelli R.L. (2001), Life in extreme environments.  \textit{Nature} 409, 1092-1101.

Sato, M., and Fukunishi H. (2005) New evidence for a link between lightning activity and tropical upper cloud coverage. \textit{Geophys. Res. Lett.} 32, L12807.

Schidlowski M. (1988) A 3,800-million-year isotopic record of life from carbon in sedimentary rocks. \textit{Nature} 333, 313-318.

Schneider, J. (2010) The Extrasolar Planets Encyclopaedia. Available online at http://exoplanet.eu.

Scott B.J. (1994) Cyclic activity in the crater lakes of Waimangu hydrothermal system, New Zealand. \textit{Geothermics} 23, 555-572.

Seager S., Schrenk M. and Bains W. (2012) An Astrophysical View of Earth-Based Metabolic Biosignature Gases.  \textit{Astrobiology} 12, 61-82.

Sekercioglu C.H., Schneider S.H., Fay J.P. and Loarie S.R. (2007) Climate Change, Elevational Range Shifts,
and Bird Extinction. \textit{Conservation Biology} 22, 140-150.

Sekiguchi M., Hayakawa M., Nickolaenko A.P. and Hobara Y. (2006) Evidence on a link between the intensity of Schumann resonance and global surface temperature.  \textit{Ann. Geophys.} 24, 1809-1817.

Smrekar S.E. Stofan E.R., Mueller N., Treiman A., Elkins-Tanton L., Helbert J., Piccioni G. and Drossart P. (2010) Recent Hotspot Volcanism on Venus from VIRTIS Emissivity Data. \textit{Science} 328, 605-608.

Spiegel D.S., Raymond S.N., Dressing C.D., Scharf C.A. and Mitchell J.L. (2010) Generalized Milankovitch Cycles and Long-Term Climatic Habitability. \textit{Astrophys. J.} 721, 1308-1318.

Strother P.K., Battison L., Brasier M.D. and Wellman C.H. (2011) Earths earliest non-marine eukaryotes. \textit{Nature} 473, 505-509.

Tarter, J.C., Backus, P.R., Mancinelli, R.L., Aurnou, J.M., Backman, D.E., Basri, G.S., Boss, A.P., Clarke, A., Deming, D., Doyle, L.R., Feigelson, E.D., Freund, F., Grinspoon, D.H., Haberle, R.J.M., Hauck, S.A., II, Heath, M.J., Henry, T.J., Hollingworth, J.L., Joshi, M.M., Kilston, S., Liu, M.C., Meikle, E., Reid, I.N., Rothschild, L.J., Scalo, J., Seguna, A., Tang, C.M., Tiedje, J.M., Turnbull, M.C., Walkowicz, L.M., Weber, A.L., and Young, R.E. (2007) A reappraisal of the habitability of planets around M Dwarf stars. \textit{Astrobiology} 7, 30-65.

Thiermann F., Akoumianaki I., Hughes J.A. and Giere O. (1997) Benthic fauna of a shallow water gaseohydrothermal vent area in the Aegean Sea (Milos, Greece).\textit{Marine Biology} 128, 149-159.

Tie X., Zhang R., Brasseur G. and Lei W. (2002) Global NOx Production by Lightning. \textit{J. Atmos. Chem} 43, 61-74.

Tomasella L., Marzari F. and Vanzani V. (1996) Evolution of the Earth obliquity after the tidal expansion of the Moon orbit. \textit{Planet. Space Sci.} 44, 427-430.

Tuttle M.D. and Stevenson D.E. (1977) Variation in the Cave Environment and its Biological Implications. National Cave Management Symposium Proceedings, 1977, Adobe Press, Albuquerque, NM, pp.108-121.

Valsami-Jones E., Baltatzis E., Bailey E.H., Boyce A.J., Alexander J.L., Magganas A., Anderson L., Waldron S. and Ragnarsdottir K.V. (2005) The geochemistry of fluids from an active shallow submarine hydrothermal system: Milos island, Hellenic Volcanic Arc. \textit{J. Volcanol. Geotherm. Res.} 148, 130-151.

V\'{a}squez M., Pall\'{e} E. and Monta\~{n}\'{e}s Rodr \'{i}guez P. (2010) The Earth as a Distant Planet, Springer Science+Business Media, New York, NY, 10013.

Walker J.C.G. (1991) Feedback Processes in the Biogeochemical Cycles of Carbon. In \textit{Scientists on Gaia}, edited by S.H. Schneider and P.J. Boston, The MIT Press, Cambridge, Massachusetts, pp.183-190.

Ward P.D. and Brownlee D. (2002) The Life and Death of Planet Earth. Times Books, New York.

Welsh W.F., Orosz J.A., Carter J.A., Fabrycky D.C., Ford E.B., Lissauer J.J., Prˇa A., Quinn S.N., Ragozzine D., Short D.R., Torres G., Winn J.N., Doyle L.R., Barclay T., Batalha N.; Bloemen S., Brugamyer E., Buchhave, L.A., Caldwell C., Caldwell D.A., Christiansen J.L., Ciardi D.R., Cochran W.D., Endl M., Fortney J.J., Gautier T.N. III, Gilliland R.L., Haas M.R., Hall J.R., Holman M.J., Howard A.W., Howell S.B., Isaacson H., Jenkins J.M., Klaus T.C., Latham D.W., Li J., Marcy G.W., Mazeh T., Quintana E.V., Robertson P., Shporer A., Steffen J.H., Windmiller G., Koch D.G. and Borucki W.J. (2012) Transiting circumbinary planets Kepler-34 b and Kepler-35 b. \textit{Nature} 481, 475-479.

Williams E., Mushtak V., Rosenfeld D., Goodman S. and Boccippio D. (2005) Thermodynamic
conditions favorable to superlative thunderstorm updraft, mixed phase microphysics and lightning flash rate. \textit{Atmos. Res.} 76, 288-306.

Williams K.E., McKay C.P., Toon O.B. and Head J.W. (2010) Do ice caves exist on Mars? \textit{Icarus} 209, 358-368.

Womack A.M., Bohannan B.J.M. and Green J.L. (2010) Biodiversity and biogeography of the atmosphere. \textit{Phil. Trans. R. Soc. B} 365, 3645-3653.

Wood C.A. (1984) Calderas: A Planetary Perspective. \textit{J. Geophys. Res.} 89, 8391-8406.

Yoder J.A., Chambers M.J., Tank J.L. and Keeney G.D. (2009) High temperature effects on water loss and survival examining the hardiness of female adults of the spider beetles, \textit{Mezium affine} and \textit{Gibbium aequinoctiale}. \textit{Journal of Insect Science} 9, Article 68

\end{document}